\documentclass[aps,pre,10pt,twocolumn,floatfix,groupedaddress,showpacs]{revtex4-1}
\usepackage[utf8]{inputenc}
\usepackage{amsmath}
\usepackage{amsfonts}
\usepackage{amssymb}
\usepackage{graphicx}
\usepackage{epstopdf}
\begin{document}

\title{Dynamic dielectric response of electrorheological fluids in drag flow}
\author{B. Horváth}
\email{bhorvath@almos.uni-pannon.hu}
\author{I. Szalai}
\email{szalai@almos.uni-pannon.hu}
\affiliation{Institute of Physics and Mechatronics, University of Pannonia, 8200 Veszprém, Hungary}

\begin{abstract}
We have determined the response time of dilute electrorheological fluids (ER) in drag flow from the dynamic dielectric response. On the basis of a kinetic rate equation a new formula was derived to approximate the experimental time-dependent dielectric permittivity during the temporal evolution of the microstructure. The dielectric response time was compared to the standard rheological response time extracted from the time-dependent shear stress, and a good agreement was obtained. We found that the dielectric method is more sensitive to detect any transient during the chain formation process. The experimental saturation value of the dielectric permittivity corresponding to the equilibrium microstructure was estimated on the basis of formulas derived from the Clausius-Mossotti equation.
\end{abstract}

\pacs{82.70.Kj, 77.22.-d, 07.50.-e}

\maketitle

\section{Introduction}
One of the main advantages of a device based on electrorheological (ER) fluids (such as a clutch, actuator, brake, or damper) is the small (of the order of $10^{-3}$~s) response time compared to conventional devices. The response time of an ER device is determined by the dynamics of the external electric field induced phase transition of the ER fluid. This reversible liquid-to-solid like state transition process involves the polarization of the dispersed particles and the microstructural change by the aggregation of the polarized particles into chainlike and columnar clusters. If the ER fluid is under mechanical deformation, then the phase transition is also influenced by the time-dependent flow of the ER fluid. The overall phase transition can be characterized by a response time, which is one of the most important parameters of an ER fluid. There are different definitions for the response time. For example, in some literature \cite{Zamudio1996} it is the time delay between the turn-on of the external electric field and the resulting change in macroscopic properties such as the rheological behavior. It can also be regarded as the time needed to form the first electrode spanning structures \cite{Liu2003}. Generally, it is defined as the time required to reach a new equilibrium microstructure after the phase transition. In this paper, we use this latter definition. The change in the macroscopic properties as a function of time during the phase transition (which is the result of the temporal evolution of the microstructure) is usually approximated with a saturating exponential function. Thus the experimental response time can be characterized with the characteristic time constant of an exponential fitting function. Besides the physical properties of the components of an ER fluid, this response time is influenced by the external conditions (electric field strength, rate of deformation, etc.).

During the operation of a typical ER device the fluid is in drag flow, in pressure-driven flow, or in a combination of the two. Because the response time is influenced by the properties of the flow, it is measured mostly in the case where the fluid is under mechanical deformation. These measurements are based on the determination of time-dependent stress \cite{Choi2009,Tian2004,Tian2005a} or pressure change \cite{Abu-Jdayil2013,Ulrich2009} of the fluid. The stress-response-based measurements are done mainly in shear mode with different geometries, while there are only a few compressive stress studies regarding the response time \cite{Tian2005b}. The microstructural change can also be investigated with optical (transmittance, light scattering) methods, therefore, these experimental techniques are suitable for response-time measurements \cite{Zamudio1996} even in the case of quiescent ER fluids. With dielectric spectroscopy the polarization process of the individual particles (relaxation time) can be investigated, but this gives only an estimate on the lower limit of the response time \cite{Weiss1992}, because it does not includes the additional time needed for structure formation.

In a previous study \cite{Horvath2012}, we employed a dielectric method to investigate the response time of quiescent ER fluids. The change in relative permittivity caused by the structure formation \cite{Wen1998a} can be measured continuously with this technique and the response time is extracted from the time-dependent permittivity as a characteristic time constant. We used this dielectric method to measure the response time of ER fluids in drag flow in addition to the quiescent state. In the case of fluids in drag flow, both the dielectric and rheological properties of the ER fluid were measured simultaneously with the improved experimental setup and the characteristic times were determined from the dielectric and rheological data.

Besides the response-time measurements, the change in dielectric properties due to structure formation was compared to theoretical predictions. We used a simple theoretical model based on formulas derived from the Clausius-Mossotti equation to calculate the difference between the permittivity of isotropic and anisotropic ER fluids in equilibrium. The model was applied to quiescent and ER fluids under shear to evaluate the experimental dielectric results.

\section{Experiment}
\subsection{Experimental setup}
\begin{figure}
\includegraphics[width=0.95\linewidth]{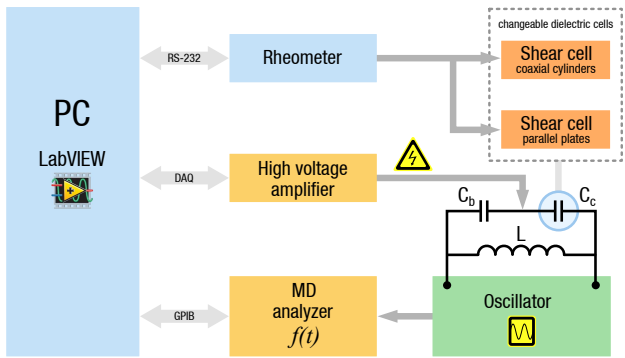}
\caption{(Color online) The block diagram of the measuring system.}
\label{fig:fig_block}
\end{figure}

The block diagram of the combined dielectric and rheological measurement setup can be seen in Fig.~\ref{fig:fig_block}. The dielectric apparatus used in our previous study was integrated with a rotational rheometer. The shear stress was measured with this rheometer at the same time as the relative permittivity of the ER fluid. This means that one HV pulse was imposed to the ER fluid and the apparatus measured both the dielectric and rheological (stress) response of the ER fluid concurrently.

The concept of the dielectric permittivity measurement is based on the determination of the frequency change of an LC oscillator \cite{Bradley1974,Brown1974}, where the capacitive element is a special, changeable dielectric cell ($\mathrm{C_c}$). There is a direct relationship between the frequency of the sinusoidally varying ac measuring field and the relative permittivity of the fluid in the dielectric cell. The measuring ac field has a frequency of $f$ = 2-3~MHz (depending on the relative permittivity of the ER fluid) and the voltage across the electrodes is $U_\mathrm{{p-p}}$ = 4~V. The strength of the electric field created by the oscillator is below the limit needed to induce structure formation in the used ER fluid. The frequency is measured by a modulation domain (MD) analyzer (HP 53310A) with high time resolution. If the ER fluid is under the influence of strong electric field ($E \approx 10^6$~V/m), then the change in relative permittivity can be continuously measured as a change in the frequency of the measuring field.
The strong electric field is created by rectangular high voltage (HV) pulses imposed to the cell. The HV pulse generator is a Trek 609E-6 HV amplifier and the analog input signal is supplied by a data acquisition card (National Instruments PCI-6052E). The slew rate of the HV amplifier is greater than 150 V/$\mu$s, therefore, even at the largest amplitude the rise time of the rectangular pulse is three orders of magnitude smaller than the measured response time of the ER fluids. The dielectric part of the measuring system was controlled with \textsc{labview} software running on a PC. The main tasks of this program were the acquisition and processing of raw data as well as to provide control functions (MD analyzer, HV pulse generator). Further technical details of the dielectric measurement setup can be found in Ref. \cite{Horvath2012}. The estimated error of the permittivity measurements is below 3\%, while the characteristic times extracted from the dielectric data have an experimental error smaller than 14\%.

\begin{figure}
\includegraphics[width=0.95\linewidth]{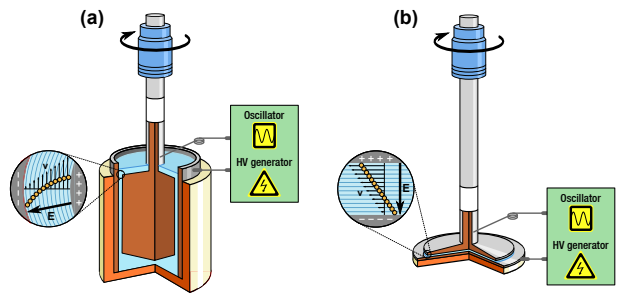}
\caption{The schematic of the dielectric cells: (a) concentric cylinders (CC) shear cell and (b) parallel plates (PP) shear cell.}
\label{fig:fig_cells}
\end{figure}

In our experiments we used two dielectric shear cells with different geometries (parallel plates and concentric cylinders) for the drag flow. The shear cells [Figs.\ \ref{fig:fig_cells}(a) and \ref{fig:fig_cells}(b)] were constructed using an Anton Paar Physica MCR301 rotational rheometer and its slightly modified accessories. The P-PTD200/E accessory and the PP50/E measuring tool is a plate-plate (PP) system (the electrode gap is 0.8~mm), while the C-PTD200/E accessory with the CC27/E tool has a concentric cylinders (CC) geometry (1.13~mm electrode gap). After small changes the measuring tools were connected as electrodes to the LC oscillator forming the frequency determining capacitive element. The electrical connection between the rotating tool and the oscillator was made by a spring loaded sliding wire. Both dielectric cells have a built-in Peltier device for temperature regulation. The drag flow in the cells can be characterized as a steady, laminar flow. The shear stress in the ER fluid was measured with the Anton Paar rheometer parallel with the dielectric measurements. The rheometer was controlled with the Anton Paar Rheoplus software, which was also used to record and process the measured data.

\subsection{Materials}
In our experiments we used dilute ER fluids containing nano-sized silica ($\mathrm{SiO_2}$) particles dispersed in silicone oil (polydimethylsiloxane) with various viscosities. The diameter of the silica particles is between 10 and 20 nm. The dynamic viscosity of the silicone oil is 0.97~$\mathrm{Pa\,s}$ at 25$^\circ{}$C. The relative permittivity of the carrier liquid is $\epsilon_\mathrm{f} = 2.7$, while the particles  have a permittivity of $\epsilon_\mathrm{p} = 4.0$. All materials were supplied by Sigma-Aldrich. Before preparing the suspensions, the silicone oil and the silica were vacuum-dried for 6 h at a pressure of 300 Pa and a temperature of 50$^\circ{}$C to remove any absorbed water. After mixing and homogenizing the two components, the suspensions were put under vacuum again for 30 min to remove the air bubbles. The volume fraction of silica particles in the ER fluids was $\phi$ = 0.04.

\section{Theory}
\subsection{Time-dependent permittivity of ER fluids}
To interpret the experimental results we employed a simple theoretical model of chain formation in ER fluids. We assume that the change in relative permittivity of the ER fluid during a HV pulse is caused by structure formation only. In a typical ER system the difference between the permittivity of the isotropic and anisotropic fluid \cite{Wen1997a,Wen1998a} is orders of magnitude higher than other nonlinear dielectric effects (e.g., electrostriction and nonlinear dielectric properties of the components), therefore the assumption can be justified.

To calculate the time-dependent permittivity of the ER fluid first we have to describe the time evolution of the microstructure. When the external electric field is switched on the particles begin to aggregate into $N$-mers (chains), where $N$ is the number of particles in one cluster. If the system is in shear flow, then the longer chains will break up into smaller fragments, and through the constant aggregation and fragmentation the microstructure will reach a new equilibrium state. In this state the length of the chains will be close to a maximum stable length ($N_{\mathrm{m}}$). Chains smaller than $N_{\mathrm{m}}$ will aggregate, while longer $N$-mers will break up.

Therefore, the time evolution of the ER system can be modeled by a phenomenological rate equation consisting of an aggregation and a fragmentation term \cite{Martin1996}:
\begin{equation}
\frac {dN(t)}{dt} = \frac {k}{N(t)} \left[ 1-\frac{N(t)^2}{N_{\mathrm{m}}^2} \right] \:.
\label{rate}
\end{equation}
The reaction rate depends on the $E$ electric field strength: $k=k_0(\epsilon_0 \epsilon_f \beta^2 E^2/8 \eta)$, where $k_0$ is a volume-fraction-dependent constant, $\epsilon_0$ is the permittivity of free space, $\eta$ is the viscosity of the carrier fluid, and $\beta=(\epsilon_\mathrm{p}-\epsilon_\mathrm{f})(\epsilon_\mathrm{p} + 2 \epsilon_\mathrm{f})$. After solving the rate equation [Eq.~\eqref{rate}] the following exact formula is obtained for the kinetics of $N$-mers:
\begin{equation}
N(t) = N_\mathrm{m} \sqrt{1-e^{-\frac{t}{\tau}} + \left(\frac{N_0}{N_\mathrm{m}}\right)^2e^{-\frac{t}{\tau}}} \:,
\label{rate_solution}
\end{equation}
where $N_0$ is the initial ($t=0$) number of particles in the $N$-mers [$N_0 = N(0)$], and $\tau$ is a characteristic time constant. In the isotropic ER fluid $N_0 \ll N_\mathrm{m}$ and Eq.~\eqref{rate_solution} can be approximated by
\begin{equation}
N(t) \simeq N_\mathrm{m} \sqrt{1-e^{-\frac{t}{\tau}}} \:.
\label{rate_solution_small}
\end{equation}
The ER fluid is treated as a two component system where the components are the single particles and the $N$-mers. The effective dielectric permittivity ($\epsilon_\mathrm{eff}$) of the ER fluid can be given by the Clausius-Mossotti equation \cite{Orihara1998}:
\begin{equation} \label{eq:C_M}
\frac{\epsilon_\mathrm{eff}-\epsilon_\mathrm{f}}{\epsilon_\mathrm{eff}+2\epsilon_\mathrm{f}} = \frac{1}{3\epsilon_0}\left( \rho_\mathrm{p}\alpha_\mathrm{p} + \rho_\mathrm{c}\alpha_\mathrm{c} \right) = y \:,
\end{equation}
where $\rho_{i}$ is the number density and $\alpha_{i}$ is the polarizability of component $i$. The subscripts p and c represent the values for single particle and chains. The base fluid of the ER system is treated as a background continuum dielectric with a relative permittivity of $\epsilon_\mathrm{f}$. The time evolution of $N$-mers results in a time-dependent effective permittivity through the $\rho(t)=N(t)/V$ number density, where $V$ is the volume of the system. Using Eq.~\eqref{eq:C_M} the effective permittivity is
\begin{equation} \label{eq:e_eff}
\epsilon_\mathrm{eff} = \epsilon_\mathrm{f} \left( \frac{1+2y}{1-y} \right) \:,
\end{equation}
which is approximated by the first two terms of its Taylor series, and because $y \ll 1$:
\begin{equation} \label{eq:e_eff_ser}
\epsilon_\mathrm{eff} \simeq \epsilon_\mathrm{f} (1 + 3y + 3y^2 + \ldots) \:.
\end{equation}
Substituting the function obtained for $N(t)$ [Eq.~\eqref{rate_solution_small}] into Eq.~\eqref{eq:C_M} and taking into account only the $N$-mers (because $\alpha_\mathrm{p} \ll \alpha_\mathrm{c}$), according to Eq.~\eqref{eq:e_eff_ser} the time-dependent permittivity can be approximated as
\begin{equation} \label{fit_bi_exp}
\Delta \epsilon(t) = \epsilon_\mathrm{eff}(t) - \epsilon_\mathrm{eff}(0) \simeq A \left( 1-e^{-\frac{t}{\tau}} \right)^\frac{1}{2} + B \left( 1-e^{-\frac{t}{\tau}} \right) \:,
\end{equation}
where $\tau$ is the characteristic time scale of the formation of longer chains, $\Delta \epsilon(t)$ is the change of permittivity in respect to the permittivity of the isotropic fluid ($t=0$), while $A$ and $B$ are constants.

According to experimental and simulation studies \cite{Jolly1999,Ly2001}, the structure formation in ER fluids progresses as two distinct processes. First, there is a fast pair formation, which is followed by the slower aggregation of the particle pairs into longer chains and columns. Therefore to take into account the pair formation, too, Eq.~\eqref{fit_bi_exp} is modified as
\begin{equation}
\begin{aligned}
\Delta \epsilon(t) =& A \left( 1-e^{-\frac{t}{\tau_1}} \right)^\frac{1}{2} + B \left( 1-e^{-\frac{t}{\tau_2}} \right)^\frac{1}{2}\\ 
&+ C \left( 1-e^{-\frac{t}{\tau_1}} \right) + D \left( 1-e^{-\frac{t}{\tau_2}} \right) \:,
\label{fit_quad_exp}
\end{aligned}
\end{equation}
where the time constant $\tau_1$ and $\tau_2$ are interpreted as the characteristic time scales for pair- and chain formation, respectively. The response time of the ER fluids is determined by these two time scales. The experimentally determined time-dependent permittivity of the ER fluids is fitted by a function according to Eq.~\eqref{fit_quad_exp}. The characteristic times of pair- and chain formation are extracted form the dielectric response to a rectangular HV pulse as the $\tau_1$ and $\tau_2$ time constants.

We remark that according to a theory by Huggins \cite{Huggins1942} describing the viscosity of dilute solution of long chain molecules a similar formula to Eq.~\eqref{fit_quad_exp} can be derived for the time-dependent viscosity of ER fluids. The $\Delta \eta$ viscosity change of dilute solution of long-chain molecules depends on the $c$ concentration as
\begin{equation} \label{eta}
\frac{\Delta \eta}{\eta_0} \simeq A c + B c^2 + \ldots \:,
\end{equation}
where $\eta_0$ is the viscosity of the pure solvent, while $A$ and $B$ are constants. Applying the theory to ER fluids the viscosity change can be approximated by Eq.~\eqref{eta}, where the time-dependent concentration of particle pairs and chains is given by Eq.~\eqref{rate_solution_small}. This yields the formula
\begin{equation}
\begin{aligned}
\Delta \eta(t) =& A' \left( 1-e^{-\frac{t}{\tau_1}} \right)^\frac{1}{2} + B' \left( 1-e^{-\frac{t}{\tau_2}} \right)^\frac{1}{2}\\ 
&+ C' \left( 1-e^{-\frac{t}{\tau_1}} \right) + D' \left( 1-e^{-\frac{t}{\tau_2}} \right) \:,
\label{fit_quad_exp_eta}
\end{aligned}
\end{equation}
which is used to approximate the experimental stress response of ER fluids instead of the $\Delta\eta(t) \simeq A'(1-e^{-t/\tau})$ equation used by other authors \cite{Tian2004,Nam2008,Wang1998}. Similarly to the dielectric response, in Eq.~\eqref{fit_quad_exp_eta} $\tau_1$ and $\tau_2$ are regarded as the time scales of pair and chain formation. In this case we call these the rheological time constants.

\subsection{The microstructure and permittivity of ER fluids}
\label{sec:perm}
The time-dependent permittivity of ER fluids tends toward a saturation value as the system reaches a new equilibrium microstructure. This can be given with the maximum of the permittivity change defined as the difference between the effective permittivities of isotropic and anisotropic ER fluids:
\begin{equation}
\Delta\epsilon_\mathrm{s}= \lim_{t\rightarrow \infty} \Delta\epsilon(t) \:.
\end{equation}
We estimate the saturation value of the permittivity change in quiescent and in ER fluids under deformation according the model outlined in the following. First, we deal with the quiescent case and then the model is modified to take into account the effect of shear.

With increasing electric field strength $\Delta \epsilon_\mathrm{s}$ is also increases, but over a threshold value of the field strength $\Delta \epsilon_\mathrm{s}$ reaches a maximum value. At higher field strengths the microstructure in quiescent fluids can be characterized as almost every particle is part of the chainlike clusters, and these structures are spanning the whole electrode gap. The permittivity of the ER fluid is calculated using the Clausius-Mossotti equation [Eq.~\eqref{eq:C_M}], but now the two components of the system are the single particles and the electrode spanning particle chains. It is assumed that the $N_\mathrm{c}$ particle long chains are one particle thick. The concentrations of the components can be given by volume fractions instead of number densities, so the $\phi_\mathrm{c}$ volume fraction of the chains is $\phi_\mathrm{c} = (1-\phi_\mathrm{p})/N_\mathrm{c}$.

The polarizability of the chains along the length is calculated using a dimensionless $h$ ''enhancement factor.'' This factor is defined as the ratio of the polarizability of the chains and the ideal additivity of the polarizability of single particles: $h=\alpha_\mathrm{c}/(N_\mathrm{c} \alpha_\mathrm{p})$. If the total number density and the polarizability is reduced by $\sigma^3$ (the diameter of the particles is $\sigma=2r$), then Eq.~(\ref{eq:C_M}) can be written as
\begin{equation} \label{eq:C_M_mix_red}
\frac{\epsilon_\mathrm{eff}-\epsilon_\mathrm{f}}{\epsilon_\mathrm{eff}+2\epsilon_\mathrm{f}} = \frac{1}{3\epsilon_0}\rho^* \left[\phi_\mathrm{p}\alpha^*_\mathrm{p} + (1-\phi_\mathrm{p})\alpha^*_\mathrm{p}h \right] \:,
\end{equation}
where the reduced quantities are marked by the superscript *. The reduced polarizability of a single particle is
\begin{equation}
\alpha^*_\mathrm{p} = \frac{1}{2} \pi \beta \epsilon_\mathrm{f} \epsilon_0 r^3 \:,
\end{equation}
where $r$ is the radius of the particle.

Kim and coworkers \cite{Kim2005} derived analytical expressions for the $h$ enhancement factor of different infinite-sized discrete clusters. The enhancement factor for infinitely long chains parallel with the electric field is
\begin{equation} \label{eq:e_factor}
h = \frac{1}{1-\dfrac{\zeta(3)}{\pi \epsilon_0}\alpha^*_\mathrm{p}} \:,
\end{equation}
where $\zeta(x)$ is the Riemann $\zeta$ function. The chains can be considered infinitely long if the $N_\mathrm{c}$ number of particles in the chains is greater than 1000. In this case the asymptotic value of $h$ of a large cluster agrees well with the value determined from the analytical expression for an infinitely long chain. In the ER fluids under investigation $N_\mathrm{c} \approx 6000$ if the chains are spanning the whole electrode gap (which is true at sufficiently large electric field strength). This means that the chains in the ER fluid can be regarded as infinitely long and Eq.~\eqref{eq:e_factor} is valid.

When the ER fluid is subjected to steady shear, the particle chains tilt due to hydrodynamic forces and are no longer parallel with the electric field. The parallel (with respect to $\textbf{E}$) component of the polarizability of the slanted chains becomes dependent on the $\theta$ tilt angle. Therefore, the formula for the enhancement factor takes the 
\begin{equation} \label{eq:e_factor_tilt}
h = \frac{1}{1-\dfrac{\zeta(3)}{2\pi \epsilon_0}(3\cos^2\theta-1)\alpha^*_\mathrm{p}}
\end{equation}
form. Equation \eqref{eq:e_factor_tilt} has a physical meaning only in the 0$^\circ < \theta <$ 54.7$^\circ$ range, because if $\theta >$~54.7$^\circ$, the interaction between the particles in the chains becomes repulsive. Substituting Eq.~\eqref{eq:e_factor_tilt} into Eq.~\eqref{eq:C_M_mix_red}, an expression is derived to calculate the effective permittivity of ER fluids under shear and in the quiescent state ($\theta$ = 0$^\circ$). In the evaluation of the experimental data we will compare the measured $\Delta\epsilon_\mathrm{s}$ with these theoretical values.

\section{Results and discussion}
\begin{figure}
\includegraphics[width=0.95\linewidth]{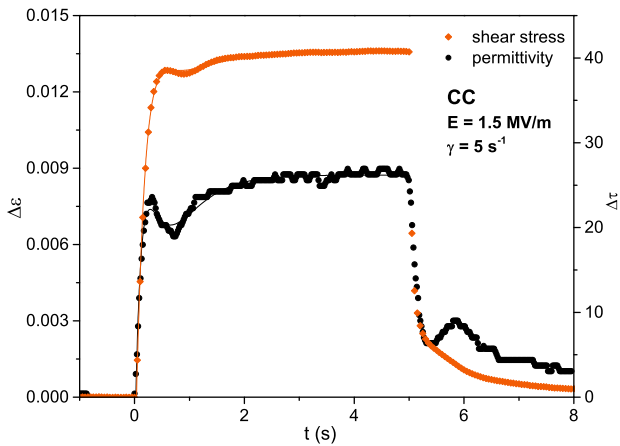}
\caption{(Color online) The concurrently measured time-dependent permittivity (black circles) and shear stress (red diamonds) of ER fluid in shear flow. The solid lines are the theoretical fitting functions according to Eqs.~\eqref{fit_quad_exp} and \eqref{fit_quad_exp_eta} between $t$~=~0-5~s. The 5~s long rectangular HV pulse was switched on at $t$~=~0~s ($\eta$ = 0.97 $\mathrm{Pa\,s}$ silicone oil, $\phi$ = 0.04, $T$ = 25$^\circ{}$C).}
\label{fig:fig_ER_e_tau}
\end{figure}

A typical concurrently measured dielectric and rheological (stress) response of the ER fluid to a rectangular HV pulse during shear is presented in Fig.~\ref{fig:fig_ER_e_tau}. The responses were determined at various electric field strengths (1.0--2.0 MV/m) and steady shear rates (0--50 $\mathrm{s}^{-1}$). In all cases the steady shear flow was applied 20~s before the electric field was switched on, and it was maintained 20~s after the HV pulse. The time-dependent permittivity was fitted with a function according to Eq.~\eqref{fit_quad_exp}, while the stress response was approximated with Eq.~\eqref{fit_quad_exp_eta}.

\subsection{Characteristic times}

\begin{figure}
\includegraphics[width=0.95\linewidth]{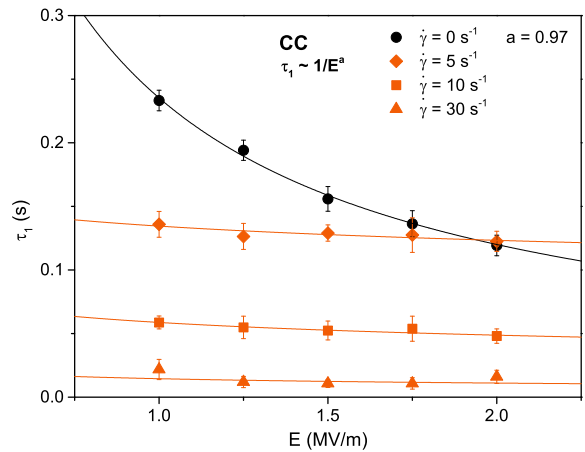}
\caption{(Color online) The $\tau_1$ characteristic time of ER fluids in drag flow (concentric cylinders) as a function of electric field strength at different shear rates ($\eta$ = 0.97 $\mathrm{Pa\,s}$ silicone oil, $\phi$ = 0.04, $T$ = 25$^\circ{}$C).}
\label{fig:fig_ER_t1_E_CC}
\end{figure}
The experimental characteristic times were determined from both the dielectric and rheological data as the time constants of the fitting functions. The following data processing was used to obtain the characteristic times. At one electric field strength and shear rate the measurements were repeated 5 times and each raw data set was fitted. The experimental characteristic times are the arithmetic mean of five values. These experimental results are presented in the following figures (Figs.~\ref{fig:fig_ER_t1_E_CC} and \ref{fig:fig_ER_PP_t1_RD_corr}), where the error bars represent the standard deviation.

The dielectric $\tau_1$ characteristic time of sheared ER fluid (CC geometry) as a function of the electric field strength can be seen in Fig.~\ref{fig:fig_ER_t1_E_CC}. At a constant shear rate, the electric field strength dependence of the experimental characteristic times can be fitted with a $\tau_1 \propto 1/E^a$ function. In the case of quiescent ($\dot{\gamma}$~=~0~$\mathrm{s}^{-1}$) ER fluids, the theoretical value of the exponent $a$ according to the polarization model of the ER effect is 2 \cite{Liu2003}. However, the experimental results show that for the ER fluid under investigation $a$ has a value approximately 1 ($a = 0.97$). The experimental response time of other ER fluids shows similar ($a <$~2) electric field strength dependence. Rejón \textit{et al.} \cite{Rejon2002} found that the exponent has a value between 0.41 and 1.05 in ER fluids containing polysaccharide particles. According to Abu-Jdayil \cite{Abu-Jdayil2013}, the response time determined by rheological measurements depends on the electric field strength with an exponent between 0.85 and 1.40 in ER fluids composed of polyurethane particles dispersed in silicone oil.

The experimental $\tau_2$ characteristic time of the formation of longer chains have a much larger error than $\tau_1$, but the following trend can be observed. In sheared ER fluids $\tau_2$ is larger by a factor of 5--10 than the corresponding $\tau_1$. This is the same trend as we found earlier \cite{Horvath2012} and was predicted by theoretical models and computer simulations \cite{Hass1993} in the case of quiescent fluids.

The experimental data show that if the ER fluid is subjected to shear, then $\tau_1$ is decreasing with increasing shear rate at a given field strength. On the other hand, at constant shear rate $\tau_1$ becomes independent of the electric field strength even at $\dot{\gamma}$ = 5~$\mathrm{s}^{-1}$. In the examined shear rate regime the exponent of the power fit is smaller than 0.39, while $a$~=~0.97$\pm$0.04 in the quiescent fluid. This behavior can be explained if we take into account the forces relevant to the structure formation in ER fluids.

The microstructure of ER fluids in steady shear flow is determined by the competition of the electrostatic forces between the particles and the viscous force. The relative importance of these forces can be described by the dimensionless Mason number \cite{BaxterDrayton1996}:
\begin{equation} \label{eq:Mn}
Mn = \frac{\eta \dot{\gamma}}{2\epsilon_\mathrm{f}\epsilon_0\beta^2 E^2} \:.
\end{equation}
In the case of the ER fluid used in our experiments at $\dot{\gamma}$ = 5~$\mathrm{s}^{-1}$ and $E$ = 2.0~MV/m the Mason number is 1.32. This means that at this shear rate the viscous forces begin to become the dominant force to determine the microstructure. If the shear rate is greater than 10~$\mathrm{s}^{-1}$, the Mason number is 2.65 $< Mn <$ 10.6 depending on the electric field strength. In this shear rate regime the viscous forces are clearly dominant, so $\tau_1$ is independent of the electric field strength.

The electric field strength dependence of the characteristic times was measured with both dielectric shear cells (parallel plates and coaxial cylinders). By the comparison of the characteristic times measured with the CC and PP cells at the same conditions, we found that the difference between $\tau_1$ is around 4--9\%. This difference stems largely from the calibration error of the parallel plates cell. The characteristic times of the studied ER fluids, therefore, can be said to be independent of the geometry of the shearing surfaces.
\begin{figure}
\includegraphics[width=0.95\linewidth]{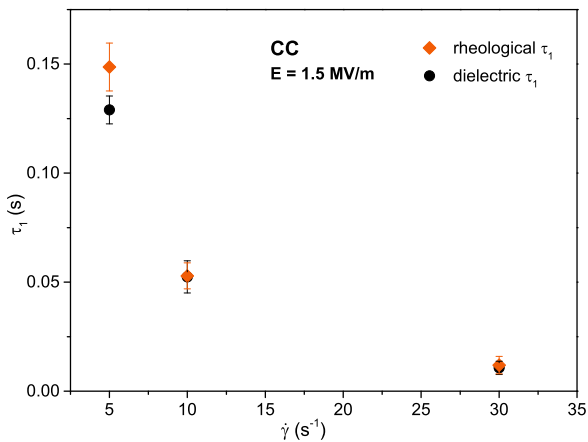}
\caption{(Color online) The characteristic times as a function of shear rate determined with the rheological (red diamonds) and dielectric (black circles) methods.}
\label{fig:fig_ER_PP_t1_RD_corr}
\end{figure}

According to our tests the measured stress response was distorted by instrumental effects, and thus the extracted characteristic times were affected. To take into account the instrumental effects we have corrected the rheological characteristic times. These corrected values agree within measurement error with the dielectric characteristic times as it can be seen in Fig.~\ref{fig:fig_ER_PP_t1_RD_corr}. One of the instrumental effects is the internal data averaging of the rheometer: The measured shear stress is averaged over half the time of the sampling interval. The internal sampling rate of the rheometer is 0.1~ms, while the measured shear stress is averaged over 25~ms, and a data point is generated at every 50~ms. Besides a time delay this internal data processing introduces low-pass filtering on the measured stress response. The estimated cutoff frequency (for --3~dB) of this filter is $f_{\mathrm{c}}\approx$ 17~Hz. Examining the measured dielectric response (where there is no averaging) in the frequency domain there are no higher frequency components in the signal. Therefore, the low-pass filtering on the stress response have no considerable effect. Besides the time delay caused by the averaging there is an additional delay which is the result of the data transfer between the instrument and the control software after each data points. In the case of the dielectric measurements no correction was necessary because the dielectric apparatus samples the permittivity without any data averaging, which means that one data point is the result of one high-precision frequency measurement. Because the raw data are stored in the memory of the MD analyzer and transferred to the control software only after the measurements there is no additional instrumental delay.

\subsection{The relationship between the permittivity and microstructure}
\begin{figure}
\includegraphics[width=0.95\linewidth]{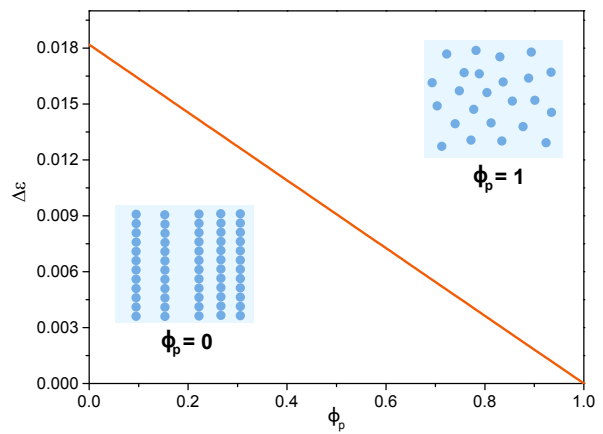}
\caption{(Color online) The theoretical effective permittivity change relative to the permittivity of the isotropic ($\phi_\mathrm{p}=1$) ER fluid as a function of the volume fraction of single particles [according to Eq.~(\ref{eq:C_M_mix_red})].}
\label{fig:fig_ER_chain_e}
\end{figure}

In the case of the quiescent ER fluids used in our experiments $\Delta \epsilon_\mathrm{s}$ reaches its maximum value if the electric field strength is $E >$~4~MV/m, indicating that almost every particle is part of the chainlike clusters. According to the experimental data the maximum of $\Delta\epsilon_\mathrm{s}$ is 0.0210$\pm$0.0005. Using the theoretical model outlined in Sec.~\ref{sec:perm}, we can give an estimate for this maximum of the permittivity increment. The theoretical effective permittivity of the ER fluid as a function of the $\phi_\mathrm{p}$ volume fraction of single particles is shown in Fig.~\ref{fig:fig_ER_chain_e}. The effective permittivity increases with decreasing particle volume fraction as the particles are aggregating into chains [see Eq.~(\ref{eq:C_M_mix_red})]. The theoretical value is $\Delta\epsilon_\mathrm{s} = 0.0182$, which agrees well with the experimental data.

\begin{figure}
\includegraphics[width=0.95\linewidth]{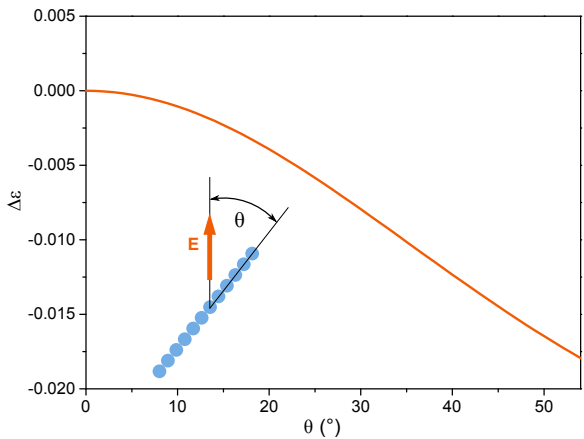}
\caption{(Color online) The theoretical effective permittivity of sheared ER fluid as the function of the $\theta$ tilt angle of slanted chains. The permittivity is given as a change relative to the permittivity of the ER fluid where the chains are parallel with the electric field ($\theta$ = 0$^\circ$).}
\label{fig:fig_ER_chain_tilt}
\end{figure}

\begin{figure}
\includegraphics[width=0.95\linewidth]{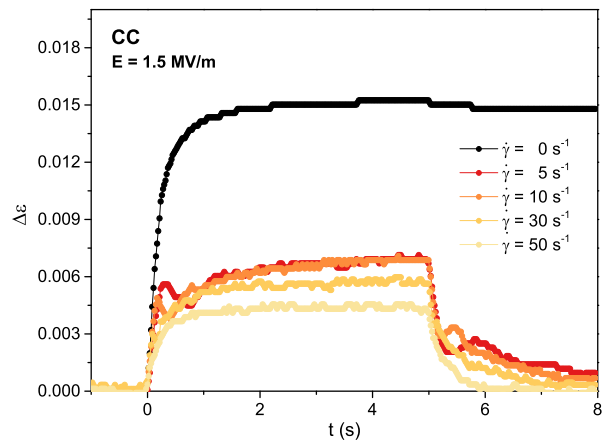}
\caption{(Color online) The dielectric response of the ER fluid in drag flow to a 5-s-long rectangular HV pulse at different shear rates ($\eta$ = 0.97 $\mathrm{Pa\,s}$ silicone oil, $\phi$ = 0.04, $T$ = 25$^\circ{}$C). The electric field was switched on at $t$~=~0~s.} 
\label{fig:fig_ER_CC_1500kVm}
\end{figure}

If the ER fluid is in shear flow, then the prediction of the model for the permittivity change as a function of the tilt angle is presented in Fig.~\ref{fig:fig_ER_chain_tilt}. As the chains become more tilted, the effective permittivity decreases. In a realistic sheared ER fluid the tilt angle of a single chain increases until the chain breaks, after which the fragments tilt less. When the fragments form a complete chain again, the tilt angle is decreasing until breakage occurs \cite{Martin1996}. This continuous break up and reforming of chains means that the slant of the chains can be described with an average angle if the microstructure reached the dynamic equilibrium state.
The measured permittivity change at various shear rates (Fig.~\ref{fig:fig_ER_CC_1500kVm}) shows a similar behavior to the theoretical prediction; as the shear rate is increased, the permittivity change becomes smaller. From the experimental permittivity data we calculated the average tilt angle using the theoretical model. If the ER fluid is subjected to a steady shear of $\dot{\gamma}$ = 5~$\mathrm{s}^{-1}$, then the measured permittivity change in respect to the permittivity of quiescent fluid is $\Delta\epsilon$ = -0.0078$\pm$0.0002. According to the employed model, this corresponds to a tilt angle of $\theta$ = 30$^\circ$, which is reasonable. At a higher $\dot{\gamma}$ = 50~$\mathrm{s}^{-1}$ shear rate $\Delta\epsilon$ = -0.0105$\pm$0.0003, which gives a value of $\theta$ = 36$^\circ$ for the average tilt angle.
Martin and Anderson \cite{Martin1996} developed a chain model of ER fluids where the chain orientation and size are found by determining the mechanical stability of a chain by balancing the hydrodynamic forces against the electrostatic forces. In the framework of this model, the critical angle when chain rupture occurs can be calculated using a rigid chain or single linear chain approximation. The theoretical estimations for the critical chain angle in the two approximations are $\theta_\mathrm{c} \approx$ 35$^\circ$ and $\theta_\mathrm{c} \approx$ 39$^\circ$. The value corresponding to the rigid chain approximation agrees well with the value of the average tilt angle calculated from our experimental data using Eqs.~(\ref{eq:C_M_mix_red}) and (\ref{eq:e_factor_tilt}).

In the dielectric response of the sheared fluid transient overshoots were observed, and these were apparent in the corresponding shear stress response, too (see Fig.~\ref{fig:fig_ER_e_tau}). The first overshoot appeared after the turn-on of the electric field, and a second was observable after the turn-off. With increasing shear rate the amplitude of the first local maximum decreased and appeared earlier (Fig.~\ref{fig:fig_ER_CC_1500kVm}). One possible explanation of this overshoot is the formation of the first long chains which lead to the permittivity and shear stress increase. It is followed by the breakage of these first chains, resulting in the local decrease of permittivity and shear stress. After breakage the fragments are reforming and the system develops toward the dynamic equilibrium of constant fragmentation and aggregation of chains. The transient after the turn-off of the electric field can be explained in a similar manner: This is the result of the last chain fragments aligning to form a complete chain before they are gradually destroyed by the shear flow. We remark that the dielectric method is more sensitive to detect these local maxima than the rheological method. Kim and coworkers \cite{Kim2004} observed similar transient overshoot in the stress response of ER fluids containing conducting polymer-coated polyethylene. They found that the phenomena could be best explained by shear-induced particle chain aggregation, where the electrical conduction of the particles also plays an important role.

\section{Conclusions}
The dielectric and stress responses of ER fluids in shear flow were characterized by measuring the response time. The dependence of the characteristic times on the electric field strength, and the shear rate were determined. The electric field strength dependence of the characteristic time was fitted with a power function, and the exponent was found to be smaller than the value predicted by the polarization model of the ER effect. We found that in the case of shear flow above a certain shear rate, the characteristic time becomes independent of the electric field strength, indicating that the viscous forces dominate the process of determining the microstructure. The time constants measured with different shearing surface geometries (parallel plates or concentric cylinders) were the same within measurement error.

We compared the characteristic time constants determined from the dielectric response of the sheared fluids to the corresponding ones determined on the basis of the shear stress behavior. The rheological characteristic times had to be corrected to take into account the instrumental response time and the effect of internal data averaging. The corrected rheological response times agreed with the dielectric response times within measurement error.

The measured change in the effective permittivity of the ER fluid was estimated on the basis of formulas derived from the Clausius-Mossotti equation. Despite the simple nature of the theoretical model, we found good agreement between the theoretical and experimental dielectric data. The model was applied to sheared ER fluids and from the measured permittivity change the average tilt angle of the chains was determined, which agrees well with the results of other authors.

\begin{acknowledgments}
We acknowledge the financial support of this work by the Hungarian State and the European Union under the TAMOP-4.2.2.A-11/1/KONV-2012-0071 project. We are thankful to Dr.-Ing. Loredana Völker-Pop at Anton Paar Germany GmbH for her help.
\end{acknowledgments}

\bibliography{references}

\begin{thebibliography}{26}%
\makeatletter
\providecommand \@ifxundefined [1]{%
 \@ifx{#1\undefined}
}%
\providecommand \@ifnum [1]{%
 \ifnum #1\expandafter \@firstoftwo
 \else \expandafter \@secondoftwo
 \fi
}%
\providecommand \@ifx [1]{%
 \ifx #1\expandafter \@firstoftwo
 \else \expandafter \@secondoftwo
 \fi
}%
\providecommand \natexlab [1]{#1}%
\providecommand \enquote  [1]{``#1''}%
\providecommand \bibnamefont  [1]{#1}%
\providecommand \bibfnamefont [1]{#1}%
\providecommand \citenamefont [1]{#1}%
\providecommand \href@noop [0]{\@secondoftwo}%
\providecommand \href [0]{\begingroup \@sanitize@url \@href}%
\providecommand \@href[1]{\@@startlink{#1}\@@href}%
\providecommand \@@href[1]{\endgroup#1\@@endlink}%
\providecommand \@sanitize@url [0]{\catcode `\\12\catcode `\$12\catcode
  `\&12\catcode `\#12\catcode `\^12\catcode `\_12\catcode `\%12\relax}%
\providecommand \@@startlink[1]{}%
\providecommand \@@endlink[0]{}%
\providecommand \url  [0]{\begingroup\@sanitize@url \@url }%
\providecommand \@url [1]{\endgroup\@href {#1}{\urlprefix }}%
\providecommand \urlprefix  [0]{URL }%
\providecommand \Eprint [0]{\href }%
\providecommand \doibase [0]{http://dx.doi.org/}%
\providecommand \selectlanguage [0]{\@gobble}%
\providecommand \bibinfo  [0]{\@secondoftwo}%
\providecommand \bibfield  [0]{\@secondoftwo}%
\providecommand \translation [1]{[#1]}%
\providecommand \BibitemOpen [0]{}%
\providecommand \bibitemStop [0]{}%
\providecommand \bibitemNoStop [0]{.\EOS\space}%
\providecommand \EOS [0]{\spacefactor3000\relax}%
\providecommand \BibitemShut  [1]{\csname bibitem#1\endcsname}%
\let\auto@bib@innerbib\@empty
%</preamble>
\bibitem [{\citenamefont {Zamudio}\ \emph {et~al.}(1996)\citenamefont
  {Zamudio}, \citenamefont {Nava}, \citenamefont {Rejón}, \citenamefont
  {Ponce}, \citenamefont {Víquez},\ and\ \citenamefont
  {Castaño}}]{Zamudio1996}%
  \BibitemOpen
  \bibfield  {author} {\bibinfo {author} {\bibfnamefont {V.}~\bibnamefont
  {Zamudio}}, \bibinfo {author} {\bibfnamefont {R.}~\bibnamefont {Nava}},
  \bibinfo {author} {\bibfnamefont {L.}~\bibnamefont {Rejón}}, \bibinfo
  {author} {\bibfnamefont {M.}~\bibnamefont {Ponce}}, \bibinfo {author}
  {\bibfnamefont {S.}~\bibnamefont {Víquez}}, \ and\ \bibinfo {author}
  {\bibfnamefont {V.}~\bibnamefont {Castaño}},\ }\href@noop {} {\bibfield
  {journal} {\bibinfo  {journal} {Physica A: Statistical Mechanics and its
  Applications}\ }\textbf {\bibinfo {volume} {227}},\ \bibinfo {pages}
  {55–65} (\bibinfo {year} {1996})}\BibitemShut {NoStop}%
\bibitem [{\citenamefont {Liu}\ \emph {et~al.}(2003)\citenamefont {Liu},
  \citenamefont {Wang},\ and\ \citenamefont {Wang}}]{Liu2003}%
  \BibitemOpen
  \bibfield  {author} {\bibinfo {author} {\bibfnamefont {Y.-L.}\ \bibnamefont
  {Liu}}, \bibinfo {author} {\bibfnamefont {B.}~\bibnamefont {Wang}}, \ and\
  \bibinfo {author} {\bibfnamefont {D.-F.}\ \bibnamefont {Wang}},\ }\href@noop
  {} {\bibfield  {journal} {\bibinfo  {journal} {Applied Mathematics and
  Mechanics}\ }\textbf {\bibinfo {volume} {24}},\ \bibinfo {pages} {385}
  (\bibinfo {year} {2003})}\BibitemShut {NoStop}%
\bibitem [{\citenamefont {Choi}\ \emph {et~al.}(2009)\citenamefont {Choi},
  \citenamefont {Nam}, \citenamefont {Yamane},\ and\ \citenamefont
  {Park}}]{Choi2009}%
  \BibitemOpen
  \bibfield  {author} {\bibinfo {author} {\bibfnamefont {B.~H.}\ \bibnamefont
  {Choi}}, \bibinfo {author} {\bibfnamefont {Y.~J.}\ \bibnamefont {Nam}},
  \bibinfo {author} {\bibfnamefont {R.}~\bibnamefont {Yamane}}, \ and\ \bibinfo
  {author} {\bibfnamefont {M.~K.}\ \bibnamefont {Park}},\ }\href@noop {}
  {\bibfield  {journal} {\bibinfo  {journal} {Journal of Physics: Conference
  Series}\ }\textbf {\bibinfo {volume} {149}},\ \bibinfo {pages} {012004}
  (\bibinfo {year} {2009})}\BibitemShut {NoStop}%
\bibitem [{\citenamefont {Tian}\ \emph {et~al.}(2004)\citenamefont {Tian},
  \citenamefont {Meng},\ and\ \citenamefont {Wen}}]{Tian2004}%
  \BibitemOpen
  \bibfield  {author} {\bibinfo {author} {\bibfnamefont {Y.}~\bibnamefont
  {Tian}}, \bibinfo {author} {\bibfnamefont {Y.}~\bibnamefont {Meng}}, \ and\
  \bibinfo {author} {\bibfnamefont {S.}~\bibnamefont {Wen}},\ }\href@noop {}
  {\bibfield  {journal} {\bibinfo  {journal} {Journal of intelligent material
  systems and structures}\ }\textbf {\bibinfo {volume} {15}},\ \bibinfo {pages}
  {621} (\bibinfo {year} {2004})}\BibitemShut {NoStop}%
\bibitem [{\citenamefont {Tian}\ \emph
  {et~al.}(2005{\natexlab{a}})\citenamefont {Tian}, \citenamefont {Li},
  \citenamefont {Zhang}, \citenamefont {Meng},\ and\ \citenamefont
  {Wen}}]{Tian2005a}%
  \BibitemOpen
  \bibfield  {author} {\bibinfo {author} {\bibfnamefont {Y.}~\bibnamefont
  {Tian}}, \bibinfo {author} {\bibfnamefont {C.}~\bibnamefont {Li}}, \bibinfo
  {author} {\bibfnamefont {M.}~\bibnamefont {Zhang}}, \bibinfo {author}
  {\bibfnamefont {Y.}~\bibnamefont {Meng}}, \ and\ \bibinfo {author}
  {\bibfnamefont {S.}~\bibnamefont {Wen}},\ }\href@noop {} {\bibfield
  {journal} {\bibinfo  {journal} {Journal of Colloid and Interface Science}\
  }\textbf {\bibinfo {volume} {288}},\ \bibinfo {pages} {290} (\bibinfo {year}
  {2005}{\natexlab{a}})}\BibitemShut {NoStop}%
\bibitem [{\citenamefont {Abu-Jdayil}(2013)}]{Abu-Jdayil2013}%
  \BibitemOpen
  \bibfield  {author} {\bibinfo {author} {\bibfnamefont {B.}~\bibnamefont
  {Abu-Jdayil}},\ }\href@noop {} {\bibfield  {journal} {\bibinfo  {journal}
  {Journal of Physics: Conference Series}\ }\textbf {\bibinfo {volume} {412}},\
  \bibinfo {pages} {012009} (\bibinfo {year} {2013})}\BibitemShut {NoStop}%
\bibitem [{\citenamefont {Ulrich}\ \emph {et~al.}(2009)\citenamefont {Ulrich},
  \citenamefont {Böhme},\ and\ \citenamefont {Bruns}}]{Ulrich2009}%
  \BibitemOpen
  \bibfield  {author} {\bibinfo {author} {\bibfnamefont {S.}~\bibnamefont
  {Ulrich}}, \bibinfo {author} {\bibfnamefont {G.}~\bibnamefont {Böhme}}, \
  and\ \bibinfo {author} {\bibfnamefont {R.}~\bibnamefont {Bruns}},\
  }\href@noop {} {\bibfield  {journal} {\bibinfo  {journal} {Journal of
  Physics: Conference Series}\ }\textbf {\bibinfo {volume} {149}},\ \bibinfo
  {pages} {012031} (\bibinfo {year} {2009})}\BibitemShut {NoStop}%
\bibitem [{\citenamefont {Tian}\ \emph
  {et~al.}(2005{\natexlab{b}})\citenamefont {Tian}, \citenamefont {Zhang},
  \citenamefont {Meng},\ and\ \citenamefont {Wen}}]{Tian2005b}%
  \BibitemOpen
  \bibfield  {author} {\bibinfo {author} {\bibfnamefont {Y.}~\bibnamefont
  {Tian}}, \bibinfo {author} {\bibfnamefont {M.}~\bibnamefont {Zhang}},
  \bibinfo {author} {\bibfnamefont {Y.}~\bibnamefont {Meng}}, \ and\ \bibinfo
  {author} {\bibfnamefont {S.}~\bibnamefont {Wen}},\ }\href@noop {} {\bibfield
  {journal} {\bibinfo  {journal} {Journal of Colloid and Interface Science}\
  }\textbf {\bibinfo {volume} {290}},\ \bibinfo {pages} {289} (\bibinfo {year}
  {2005}{\natexlab{b}})}\BibitemShut {NoStop}%
\bibitem [{\citenamefont {Weiss}\ and\ \citenamefont
  {Carlson}(1992)}]{Weiss1992}%
  \BibitemOpen
  \bibfield  {author} {\bibinfo {author} {\bibfnamefont {K.~D.}\ \bibnamefont
  {Weiss}}\ and\ \bibinfo {author} {\bibfnamefont {J.~D.}\ \bibnamefont
  {Carlson}},\ }\href@noop {} {\bibfield  {journal} {\bibinfo  {journal}
  {International Journal of Modern Physics B}\ }\textbf {\bibinfo {volume}
  {6}},\ \bibinfo {pages} {2609} (\bibinfo {year} {1992})}\BibitemShut
  {NoStop}%
\bibitem [{\citenamefont {Horváth}\ and\ \citenamefont
  {Szalai}(2012)}]{Horvath2012}%
  \BibitemOpen
  \bibfield  {author} {\bibinfo {author} {\bibfnamefont {B.}~\bibnamefont
  {Horváth}}\ and\ \bibinfo {author} {\bibfnamefont {I.}~\bibnamefont
  {Szalai}},\ }\href@noop {} {\bibfield  {journal} {\bibinfo  {journal}
  {Physical Review E}\ }\textbf {\bibinfo {volume} {86}},\ \bibinfo {pages}
  {061403} (\bibinfo {year} {2012})}\BibitemShut {NoStop}%
\bibitem [{\citenamefont {Wen}\ \emph {et~al.}(1998)\citenamefont {Wen},
  \citenamefont {Ma}, \citenamefont {Tam},\ and\ \citenamefont
  {Sheng}}]{Wen1998a}%
  \BibitemOpen
  \bibfield  {author} {\bibinfo {author} {\bibfnamefont {W.}~\bibnamefont
  {Wen}}, \bibinfo {author} {\bibfnamefont {H.}~\bibnamefont {Ma}}, \bibinfo
  {author} {\bibfnamefont {W.~Y.}\ \bibnamefont {Tam}}, \ and\ \bibinfo
  {author} {\bibfnamefont {P.}~\bibnamefont {Sheng}},\ }\href@noop {}
  {\bibfield  {journal} {\bibinfo  {journal} {Applied Physics Letters}\
  }\textbf {\bibinfo {volume} {73}},\ \bibinfo {pages} {3070} (\bibinfo {year}
  {1998})}\BibitemShut {NoStop}%
\bibitem [{\citenamefont {Bradley}\ and\ \citenamefont
  {Jones}(1974)}]{Bradley1974}%
  \BibitemOpen
  \bibfield  {author} {\bibinfo {author} {\bibfnamefont {P.}~\bibnamefont
  {Bradley}}\ and\ \bibinfo {author} {\bibfnamefont {G.}~\bibnamefont
  {Jones}},\ }\href@noop {} {\bibfield  {journal} {\bibinfo  {journal} {Journal
  of Physics E: Scientific Instruments}\ }\textbf {\bibinfo {volume} {7}},\
  \bibinfo {pages} {449} (\bibinfo {year} {1974})}\BibitemShut {NoStop}%
\bibitem [{\citenamefont {Brown}\ \emph {et~al.}(1974)\citenamefont {Brown},
  \citenamefont {Jones},\ and\ \citenamefont {Davies}}]{Brown1974}%
  \BibitemOpen
  \bibfield  {author} {\bibinfo {author} {\bibfnamefont {B.}~\bibnamefont
  {Brown}}, \bibinfo {author} {\bibfnamefont {G.}~\bibnamefont {Jones}}, \ and\
  \bibinfo {author} {\bibfnamefont {M.}~\bibnamefont {Davies}},\ }\href@noop {}
  {\bibfield  {journal} {\bibinfo  {journal} {Journal of Physics D: Applied
  Physics}\ }\textbf {\bibinfo {volume} {7}},\ \bibinfo {pages} {1192}
  (\bibinfo {year} {1974})}\BibitemShut {NoStop}%
\bibitem [{\citenamefont {Wen}\ \emph {et~al.}(1997)\citenamefont {Wen},
  \citenamefont {Men},\ and\ \citenamefont {Lu}}]{Wen1997a}%
  \BibitemOpen
  \bibfield  {author} {\bibinfo {author} {\bibfnamefont {W.}~\bibnamefont
  {Wen}}, \bibinfo {author} {\bibfnamefont {S.}~\bibnamefont {Men}}, \ and\
  \bibinfo {author} {\bibfnamefont {K.}~\bibnamefont {Lu}},\ }\href@noop {}
  {\bibfield  {journal} {\bibinfo  {journal} {Physical Review E}\ }\textbf
  {\bibinfo {volume} {55}},\ \bibinfo {pages} {3015} (\bibinfo {year}
  {1997})}\BibitemShut {NoStop}%
\bibitem [{\citenamefont {Martin}\ and\ \citenamefont
  {Anderson}(1996)}]{Martin1996}%
  \BibitemOpen
  \bibfield  {author} {\bibinfo {author} {\bibfnamefont {J.~E.}\ \bibnamefont
  {Martin}}\ and\ \bibinfo {author} {\bibfnamefont {R.~A.}\ \bibnamefont
  {Anderson}},\ }\href@noop {} {\bibfield  {journal} {\bibinfo  {journal}
  {Journal of Chemical Physics}\ }\textbf {\bibinfo {volume} {104}},\ \bibinfo
  {pages} {4814} (\bibinfo {year} {1996})}\BibitemShut {NoStop}%
\bibitem [{\citenamefont {Orihara}\ \emph {et~al.}(1998)\citenamefont
  {Orihara}, \citenamefont {Sugaya}, \citenamefont {Tajiri}, \citenamefont
  {Ishibashi}, \citenamefont {Doi},\ and\ \citenamefont {Inoue}}]{Orihara1998}%
  \BibitemOpen
  \bibfield  {author} {\bibinfo {author} {\bibfnamefont {H.}~\bibnamefont
  {Orihara}}, \bibinfo {author} {\bibfnamefont {M.}~\bibnamefont {Sugaya}},
  \bibinfo {author} {\bibfnamefont {K.}~\bibnamefont {Tajiri}}, \bibinfo
  {author} {\bibfnamefont {Y.}~\bibnamefont {Ishibashi}}, \bibinfo {author}
  {\bibfnamefont {M.}~\bibnamefont {Doi}}, \ and\ \bibinfo {author}
  {\bibfnamefont {A.}~\bibnamefont {Inoue}},\ }\href@noop {} {\bibfield
  {journal} {\bibinfo  {journal} {Journal of the Physical Society of Japan}\
  }\textbf {\bibinfo {volume} {67}},\ \bibinfo {pages} {655} (\bibinfo {year}
  {1998})}\BibitemShut {NoStop}%
\bibitem [{\citenamefont {Jolly}\ \emph {et~al.}(1999)\citenamefont {Jolly},
  \citenamefont {Bender},\ and\ \citenamefont {Mathers}}]{Jolly1999}%
  \BibitemOpen
  \bibfield  {author} {\bibinfo {author} {\bibfnamefont {M.~R.}\ \bibnamefont
  {Jolly}}, \bibinfo {author} {\bibfnamefont {J.~W.}\ \bibnamefont {Bender}}, \
  and\ \bibinfo {author} {\bibfnamefont {R.~T.}\ \bibnamefont {Mathers}},\
  }\href@noop {} {\bibfield  {journal} {\bibinfo  {journal} {International
  Journal of Modern Physics B}\ }\textbf {\bibinfo {volume} {13}},\ \bibinfo
  {pages} {2036} (\bibinfo {year} {1999})}\BibitemShut {NoStop}%
\bibitem [{\citenamefont {Ly}\ \emph {et~al.}(2001)\citenamefont {Ly},
  \citenamefont {Ito}, \citenamefont {Banks}, \citenamefont {Jolly},\ and\
  \citenamefont {Reitich}}]{Ly2001}%
  \BibitemOpen
  \bibfield  {author} {\bibinfo {author} {\bibfnamefont {H.}~\bibnamefont
  {Ly}}, \bibinfo {author} {\bibfnamefont {K.}~\bibnamefont {Ito}}, \bibinfo
  {author} {\bibfnamefont {H.}~\bibnamefont {Banks}}, \bibinfo {author}
  {\bibfnamefont {M.}~\bibnamefont {Jolly}}, \ and\ \bibinfo {author}
  {\bibfnamefont {F.}~\bibnamefont {Reitich}},\ }\href@noop {} {\bibfield
  {journal} {\bibinfo  {journal} {International Journal of Modern Physics B}\
  }\textbf {\bibinfo {volume} {15}},\ \bibinfo {pages} {894} (\bibinfo {year}
  {2001})}\BibitemShut {NoStop}%
\bibitem [{\citenamefont {Huggins}(1942)}]{Huggins1942}%
  \BibitemOpen
  \bibfield  {author} {\bibinfo {author} {\bibfnamefont {M.~L.}\ \bibnamefont
  {Huggins}},\ }\href@noop {} {\bibfield  {journal} {\bibinfo  {journal}
  {Journal of the American Chemical Society}\ }\textbf {\bibinfo {volume}
  {64}},\ \bibinfo {pages} {2716} (\bibinfo {year} {1942})}\BibitemShut
  {NoStop}%
\bibitem [{\citenamefont {Nam}\ \emph {et~al.}(2008)\citenamefont {Nam},
  \citenamefont {Park},\ and\ \citenamefont {Yamane}}]{Nam2008}%
  \BibitemOpen
  \bibfield  {author} {\bibinfo {author} {\bibfnamefont {Y.}~\bibnamefont
  {Nam}}, \bibinfo {author} {\bibfnamefont {M.}~\bibnamefont {Park}}, \ and\
  \bibinfo {author} {\bibfnamefont {R.}~\bibnamefont {Yamane}},\ }\href@noop {}
  {\bibfield  {journal} {\bibinfo  {journal} {Experiments in Fluids}\ }\textbf
  {\bibinfo {volume} {44}},\ \bibinfo {pages} {915} (\bibinfo {year}
  {2008})}\BibitemShut {NoStop}%
\bibitem [{\citenamefont {Wang}\ \emph {et~al.}(1998)\citenamefont {Wang},
  \citenamefont {Lin}, \citenamefont {Fang},\ and\ \citenamefont
  {Tao}}]{Wang1998}%
  \BibitemOpen
  \bibfield  {author} {\bibinfo {author} {\bibfnamefont {Z.}~\bibnamefont
  {Wang}}, \bibinfo {author} {\bibfnamefont {Z.}~\bibnamefont {Lin}}, \bibinfo
  {author} {\bibfnamefont {H.}~\bibnamefont {Fang}}, \ and\ \bibinfo {author}
  {\bibfnamefont {R.}~\bibnamefont {Tao}},\ }\href@noop {} {\bibfield
  {journal} {\bibinfo  {journal} {Journal of Applied Physics}\ }\textbf
  {\bibinfo {volume} {83}},\ \bibinfo {pages} {1125} (\bibinfo {year}
  {1998})}\BibitemShut {NoStop}%
\bibitem [{\citenamefont {Kim}\ \emph {et~al.}(2005)\citenamefont {Kim},
  \citenamefont {Sofo}, \citenamefont {Velegol}, \citenamefont {Cole},\ and\
  \citenamefont {Mukhopadhyay}}]{Kim2005}%
  \BibitemOpen
  \bibfield  {author} {\bibinfo {author} {\bibfnamefont {H.-Y.}\ \bibnamefont
  {Kim}}, \bibinfo {author} {\bibfnamefont {J.~O.}\ \bibnamefont {Sofo}},
  \bibinfo {author} {\bibfnamefont {D.}~\bibnamefont {Velegol}}, \bibinfo
  {author} {\bibfnamefont {M.~W.}\ \bibnamefont {Cole}}, \ and\ \bibinfo
  {author} {\bibfnamefont {G.}~\bibnamefont {Mukhopadhyay}},\ }\href@noop {}
  {\bibfield  {journal} {\bibinfo  {journal} {Physical Review A}\ }\textbf
  {\bibinfo {volume} {72}},\ \bibinfo {pages} {053201} (\bibinfo {year}
  {2005})}\BibitemShut {NoStop}%
\bibitem [{\citenamefont {Rejón}\ \emph {et~al.}(2002)\citenamefont {Rejón},
  \citenamefont {Ramírez}, \citenamefont {Paz}, \citenamefont {Goycoolea},\
  and\ \citenamefont {Valdez}}]{Rejon2002}%
  \BibitemOpen
  \bibfield  {author} {\bibinfo {author} {\bibfnamefont {L.}~\bibnamefont
  {Rejón}}, \bibinfo {author} {\bibfnamefont {A.}~\bibnamefont {Ramírez}},
  \bibinfo {author} {\bibfnamefont {F.}~\bibnamefont {Paz}}, \bibinfo {author}
  {\bibfnamefont {F.}~\bibnamefont {Goycoolea}}, \ and\ \bibinfo {author}
  {\bibfnamefont {M.}~\bibnamefont {Valdez}},\ }\href@noop {} {\bibfield
  {journal} {\bibinfo  {journal} {Carbohydrate Polymers}\ }\textbf {\bibinfo
  {volume} {48}},\ \bibinfo {pages} {413} (\bibinfo {year} {2002})}\BibitemShut
  {NoStop}%
\bibitem [{\citenamefont {Hass}(1993)}]{Hass1993}%
  \BibitemOpen
  \bibfield  {author} {\bibinfo {author} {\bibfnamefont {K.}~\bibnamefont
  {Hass}},\ }\href@noop {} {\bibfield  {journal} {\bibinfo  {journal} {Physical
  Review E}\ }\textbf {\bibinfo {volume} {47}},\ \bibinfo {pages} {3362}
  (\bibinfo {year} {1993})}\BibitemShut {NoStop}%
\bibitem [{\citenamefont {Baxter-Drayton}\ and\ \citenamefont
  {Brady}(1996)}]{BaxterDrayton1996}%
  \BibitemOpen
  \bibfield  {author} {\bibinfo {author} {\bibfnamefont {Y.}~\bibnamefont
  {Baxter-Drayton}}\ and\ \bibinfo {author} {\bibfnamefont {J.~F.}\
  \bibnamefont {Brady}},\ }\href@noop {} {\bibfield  {journal} {\bibinfo
  {journal} {Journal of Rheology}\ }\textbf {\bibinfo {volume} {40}},\ \bibinfo
  {pages} {1027} (\bibinfo {year} {1996})}\BibitemShut {NoStop}%
\bibitem [{\citenamefont {Kim}\ and\ \citenamefont {Park}(2004)}]{Kim2004}%
  \BibitemOpen
  \bibfield  {author} {\bibinfo {author} {\bibfnamefont {Y.~D.}\ \bibnamefont
  {Kim}}\ and\ \bibinfo {author} {\bibfnamefont {D.~H.}\ \bibnamefont {Park}},\
  }\href@noop {} {\bibfield  {journal} {\bibinfo  {journal} {Synthetic Metals}\
  }\textbf {\bibinfo {volume} {142}},\ \bibinfo {pages} {147} (\bibinfo {year}
  {2004})}\BibitemShut {NoStop}%
\end{thebibliography}%

\end{document}